\documentstyle[12pt]{article}

\catcode`\@=11
\long\def\@makefntext#1{
\protect\noindent \hbox to 3.2pt {\hskip-.9pt
$^{{\ninerm\@thefnmark}}$\hfil}#1\hfill}		

\def\@makefnmark{\hbox to 0pt{$^{\@thefnmark}$\hss}}  

\def\ps@myheadings{\let\@mkboth\@gobbletwo
\def\@oddhead{\hbox{}
\rightmark\hfil\ninerm\thepage}
\def\@oddfoot{}\def\@evenhead{\ninerm\thepage\hfil
\leftmark\hbox{}}\def\@evenfoot{}
\def\sectionmark##1{}\def\subsectionmark##1{}}
\newcommand{\bra}[1]{\langle #1 |}
\newcommand{\ket}[1]{| #1 \rangle}

\def\const{\mbox{const}}
\def\e{\mbox{e}}

\renewcommand{\thefootnote}{\fnsymbol{footnote}}

\newcounter{sectionc}\newcounter{subsectionc}\newcounter{subsubsectionc}
\renewcommand{\section}[1] {\vspace*{0.6cm}\addtocounter{sectionc}{1}
\setcounter{subsectionc}{0}\setcounter{subsubsectionc}{0}\noindent
	{\normalsize\bf\thesectionc. #1}\par\vspace*{0.4cm}}
\renewcommand{\subsection}[1] {\vspace*{0.6cm}\addtocounter{subsectionc}{1}
	\setcounter{subsubsectionc}{0}\noindent
	{\normalsize\it\thesectionc.\thesubsectionc. #1}\par\vspace*{0.4cm}}
\renewcommand{\subsubsection}[1]
{\vspace*{0.6cm}\addtocounter{subsubsectionc}{1}
	\noindent {\normalsize\rm\thesectionc.\thesubsectionc.\thesubsubsectionc.
	#1}\par\vspace*{0.4cm}}

\newcounter{appendixc}
\newcounter{subappendixc}[appendixc]
\newcounter{subsubappendixc}[subappendixc]

\renewcommand{\appendix}[1] {\vspace*{0.6cm}
        \refstepcounter{appendixc}
        \setcounter{figure}{0}
        \setcounter{table}{0}
        \setcounter{equation}{0}
        \renewcommand{\thefigure}{\Alph{appendixc}.\arabic{figure}}
        \renewcommand{\thetable}{\Alph{appendixc}.\arabic{table}}
        \renewcommand{\theappendixc}{\Alph{appendixc}}
        \renewcommand{\theequation}{\Alph{appendixc}.\arabic{equation}}
        \noindent{\bf Appendix \theappendixc #1}\par\vspace*{0.4cm}}

\def\abstracts#1{{

\centering{\begin{minipage}{12.2truecm}\footnotesize\baselineskip=12pt\noindent
	\centerline{\footnotesize ABSTRACT}\vspace*{0.3cm}
	\parindent=0pt #1
	\end{minipage}}\par}}

\renewenvironment{thebibliography}[1]
	{\begin{list}{\arabic{enumi}.}
	{\usecounter{enumi}\setlength{\parsep}{0pt}
\setlength{\leftmargin 1.25cm}{\rightmargin 0pt}
\setlength{\leftmargin 0.52cm}{\rightmargin 0pt}
	 \setlength{\itemsep}{0pt} \settowidth
	{\labelwidth}{#1.}\sloppy}}{\end{list}}

\newcounter{itemlistc}
\newcounter{romanlistc}
\newcounter{alphlistc}
\newcounter{arabiclistc}

\newcommand{\fcaption}[1]{
        \refstepcounter{figure}
        \setbox\@tempboxa = \hbox{\footnotesize Fig.~\thefigure. #1}
        \ifdim \wd\@tempboxa > 6in
           {\begin{center}
        \parbox{6in}{\footnotesize\baselineskip=12pt Fig.~\thefigure. #1}
            \end{center}}
        \else
             {\begin{center}
             {\footnotesize Fig.~\thefigure. #1}
              \end{center}}
        \fi}

\newcommand{\tcaption}[1]{
        \refstepcounter{table}
        \setbox\@tempboxa = \hbox{\footnotesize Table~\thetable. #1}
        \ifdim \wd\@tempboxa > 6in
           {\begin{center}
        \parbox{6in}{\footnotesize\baselineskip=12pt Table~\thetable. #1}
            \end{center}}
        \else
             {\begin{center}
             {\footnotesize Table~\thetable. #1}
              \end{center}}
        \fi}

\def\@citex[#1]#2{\if@filesw\immediate\write\@auxout
	{\string\citation{#2}}\fi
\def\@citea{}\@cite{\@for\@citeb:=#2\do
	{\@citea\def\@citea{,}\@ifundefined
	{b@\@citeb}{{\bf ?}\@warning
	{Citation `\@citeb' on page \thepage \space undefined}}
	{\csname b@\@citeb\endcsname}}}{#1}}

\newif\if@cghi
\def\cite{\@cghitrue\@ifnextchar [{\@tempswatrue
	\@citex}{\@tempswafalse\@citex[]}}
\def\citelow{\@cghifalse\@ifnextchar [{\@tempswatrue
	\@citex}{\@tempswafalse\@citex[]}}
\def\@cite#1#2{{$\null^{#1}$\if@tempswa\typeout
	{IJCGA warning: optional citation argument
	ignored: `#2'} \fi}}

 1
 1
 1

\font\ninerm=cmr9

\begin{document}
\vspace*{3.6cm}

\centerline{\normalsize\bf NON-PERTURBATIVE ASPECTS OF
              MULTIPARTICLE PRODUCTION}
\vspace*{1cm}
\centerline{\footnotesize V.A.RUBAKOV}
\baselineskip=13pt
\centerline{\footnotesize\it Institute for Nuclear Research of the Russian
Academy of Sciences, 60th October Prospect 7a, }
\baselineskip=12pt
\centerline{\footnotesize\it Moscow, 117312, Russia}
\centerline{\footnotesize E-mail: rubakov@ms2.inr.ac.ru}
\centerline{\it hep-ph/9511236}
\vspace*{3cm}
\abstracts{ Processes with multiparticle final states
in weakly coupled theories, both with
and without instantons, cannot be studied
perturbatively at most interesting energies and multiplicities.
Semiclassical approaches to the calculation of their
total cross sections are reviewed.}

\vspace*{0.6cm}
\normalsize\baselineskip=15pt
\setcounter{footnote}{0}
\renewcommand{\thefootnote}{\alph{footnote}}
\section{Introduction}

In this paper I discuss interacting bosonic theories (fermions are
believed to play  minor role in the processes to be considered) whose
coupling constant, generically denoted by $\alpha$, is small. At c.m.
energies of order $m/\alpha$, where $m$ is the typical mass of elementary
quanta, the number of particles kinematically allowed to be produced is of
order $1/\alpha$. It is this multiplicity that will be of interest in this
paper. The non-perturbative character of the the processes where the
number of outgoing particles is so large may be seen already from
combinatorics of the corresponding graphs. On the other hand, energies of
interst are not exponentially high, $\alpha \ln \frac{E}{m} \ll 1$,
therefore, such
aspects as running of the coupling constant or triviality are irrelevant.

There are two classes of multi-particle processes that are discussed most
widely. Historically, considerable interest has been attracted by {\it
instanton-like} processes \cite{Ringwald,Espinosa} which, in the
electroweak theory, violate baryon and lepton numbers \cite{tHooft}.
It has been realized \cite{Cornwall,Goldberg} that multiparticle processes
{\it without instantons} show remarkably similar behavior. From purely
theoretical point of view, both classes of processes are of interest
because leading order calculations indicate rapid increase of the cross
sections at high energies and multiplicities, the "corrections" are more
than important at energies and multiplicities of order $1/\alpha$ and,
ultimately, novel techniques of semiclassical type should be developed for
the complete treatment of the cross sections. On the phenomenological side,
the processes under study include multiple $W$- and Higgs boson production
(with or without violation of $B$ and $L$) in the electroweak theory at
energies in multi-TeV range and also possibly multiple production of
semi-hard gluons in QCD; clearly, these phenomena, if occur at reasonably
high rates, are of general interest.

\section{Perturbative Results}
Let me briefly summarize perturbative results on the multi-particle
processes; in the case of instanton-like processes, "perturbation theory"
means perturbation theory about an instanton. I will not discuss most of
perturbative techniques, even though they are of interest by themselves
and provide some important hints for non-perturbative analysis; for reviews
see refs. \cite{Mattis,TinyakovIJMPA,VoloshinReview}.

\subsection{Instanton-Like Processes}

As a prototype model for  instanton-like processes at high
energies,  consider an $SU(2)$ gauge theory with one Higgs doublet
(minimal standard model in the limit $\sin^{2}\theta_{W} = 0$). In the
leading order of perturbation theory about an instanton, the cross section
of the production of a given number of $W$- and Higgs bosons rapidly
increases with c.m. energy $E$ \cite{Ringwald,Espinosa}. The total
instanton-like cross section in the leading order shows the exponential
behavior \cite{MVV,Zakharov1,KRT1,Porratti},
\begin{equation}
\sigma_{inst}^{leading~order}\propto
      \exp\left[\frac{4\pi}{\alpha_{W}}\left(-1
      + \frac{9}{8}\left(\frac{E}{E_0}\right)^{4/3}\right)\right]
\label{5*}
\end{equation}
where $E_{0}=\sqrt{6}\pi m_{W}/\alpha_{W}$ is of order 15 TeV in the
electroweak theory. The number of $W$-bosons produced is of order
\[
 n\sim \frac{4\pi}{\alpha_{W}}
      \left(\frac{E}{E_0}\right)^{4/3}
\]
The cross section (\ref{5*}), being exponentially small at low energy,
exponentially increases with energy and hits the unitarity bound at
$E\sim E_{0}$; at these energies the multiplicity of outgoing particles
becomes of order $n\sim 1/\alpha_{W}$.

The expression (\ref{5*}) should not determine the true cross section at
all energies, as it violates unitarity. Indeed, corrections to the leading
order result (\ref{5*}) become large at $E\sim E_{0}$. Generally, the total
cross section has the following functional form
\cite{KRT1,Yaffe,ArnoldMattis}
\begin{equation}
    \sigma_{inst}\propto
      \exp\left[\frac{4\pi}{\alpha_{W}}
      F\left(\frac{E}{E_0}\right)\right]
\label{5+}
\end{equation}
and perturbation theory about the instanton provides the expansion
\cite{Khoze,ArnMatt,DiakPet,Mueller1}
of $F(E/E_{0})$ in powers of $(E/E_{0})^{2/3}$,
\[
 F\left(\frac{E}{E_{0}}\right) = -1 +
        \frac{9}{8}\left(\frac{E}{E_{0}}\right)^{\frac{4}{3}}
       - \frac{9}{16}\left(\frac{E}{E_{0}}\right)^{\frac{6}{3}}
       + \dots
\]
This series blows up at $E\sim E_{0}$, and new non-perturbative
techniques are needed to find the exponent of the cross section, $F$,
in the interesting energy range.

A few remarks are in order. First, the exponent
$F(E/E_{0})$
in Eq. (\ref{5+})
receives contributions both from tree graphs (about an instanton)
and from {\it loop} graphs; the latter involve hard incoming particles
\cite{Mueller2,Mueller3}
 with energies of order $E$. Second, Eq. (\ref{5+})
looks semiclassically, and one naturally expects that there exisis
semiclassical technique for calculating the exponent. This
semiclassical
technique should incorporate loop effects about an instanton (!).
Third, the behavior
of
$F(E/E_{0})$ (whether it becomes zero at some energy or remains always
negative) determines whether the total instanton cross section becomes
unsuppressed at some energy or is exponentially small at all energies.
Finally, the general form of Eq. (\ref{5+}) is valid in all weakly coupled
models with instantons, though the perturbative series for
$F(E/E_{0})$ is model-dependent.

\subsection{Processes without Instantons}

As a prototype model with multi-particle production and trivial vacuum
consider massive $\frac{\lambda}{4} \phi^{4}$ theory without spontaneous
breaking of the symmetry $\phi \to -\phi$. The direct analysis of {\it tree}
diagrams with $n$ outgoing particles is possible at exact threshold, i.e.,
when all final particles are at rest. The exact result for the tree
amplitude at threshold is \cite{VoloshinNP}
\begin{equation}
    A^{tree}_{1\to n} (E=nm) =
		 n! \left(\frac{\lambda}{8m^2}\right)^{\frac{n-1}{2}}
\label{7*}
\end{equation}
The factorial behavior of the amplitudes indicates that the tree cross
sections also increase with $n$ (if amplitudes do not drop too rapidly with
energy),
\begin{equation}
    \sigma^{tree}_{1\to n} \sim
		 \frac{1}{n!}
   | A^{tree}_{1\to n}|^{2}\times (phase~ space)
	      \sim n! \lambda^{n} \epsilon^{n}
\label{7**}
\end{equation}
where
\[
\epsilon = (E -nm)/n
\]
 is the average kinetic energy of outgoing
particles. Indeed, the direct analysis of the tree diagrams
\cite{VoloshinEstimate,AKPEstimate} has lead to a lower bound for the tree
cross section which grows factorially with $n$ in a way similar to Eq.
(\ref{7**}).

At $n \sim 1/\lambda$ the tree cross section (\ref{7**}) grows with $n$
and exceeds the unitarity limit at large enough $n$; the corresponding
total c.m. energy is of order $m/\lambda$. This means that loop
"corrections" must be large at these multiplicities. In fact, the one loop
contribution
\cite{VoloshinOneloop}
 to the tree amplitude at thershold becomes large even before
the multiplicity becomes of order $1/\lambda$,
\begin{equation}
    A^{tree}_{1\to n} +
    A^{one loop}_{1\to n} =
    A^{tree}_{1\to n} (1 + B\lambda n^2)
 \label{8+}
\end{equation}
where (complex) numerical coefficient $B$ is of order one and has been
calculated in refs. \cite{VoloshinOneloop,AKPOneLoop}. Further perturbative
analysis\footnote{In fact, the results (\ref{8+}), (\ref{8*}) and
(\ref{8**}) have been obtained by making use of the functional technique of
ref.  \cite{Brown}. I will discuss this technique in some detail later.}
 has shown \cite{LRST} that both the loop corrections and energy corrections
exponentiate, which provides a strong argument showing that the true cross
section has exponential behavior,
\begin{equation}
  \sigma_{1\to n}(E) \propto \exp\left[n F(\lambda n, \epsilon)\right]
\label{8*}
\end{equation}
in the regime of small $\lambda$, large $n$ and
$\lambda n,\epsilon =$ fixed. The loop expansion for the exponent $F$ is
the expansion in $(\lambda n)$; only a few terms in the expansion of $F$ at
small $\lambda n$ and $\epsilon$ are known,
\begin{equation}
     F(\lambda n, \epsilon) =
	    \ln \frac{\lambda n}{16} + \frac{1}{2}
	     +\frac{3}{2}\ln \frac{\epsilon}{3\pi} -
	     \frac{17}{12}\epsilon
	     + B\lambda n + \dots
\label{8**}
\end{equation}
In complete analogy to the instanton case, the cross section (\ref{8*})
shows the semiclassical behavior, and the perturbation theory for the
exponent $F$ blows up in the most interesting region $\lambda n \sim 1$
(the energy dependence of the tree contribution into the
exponent is by now a technical matter, as I
will discuss below).

Thus, the perturbative calculations of processes with multiparticle final
states (both in the presence and without instantons) provide important
insight into the functional form of the cross sections. However, these
calculations are reliable only at relatively small $n$ when the cross
sections are still exponentially suppressed. Attacking the most challeging
problem of really large $n$ requires the development of novel
non-perturbative techniques which, as suggested by Eqs. (\ref{5+}) and
(\ref{8*}), should be of  semiclassical type. Before turning to some of the
attempts in this direction, let me make a  comment concerning the
expectations for the behavior of multiparticle cross sections, which is
based on rather general grounds.

\subsection{Unitarity}

The argument I would like to present is in the spirit of refs.
\cite{ZakharovUnitar,Veneziano} and is based on unitarity and ordinary
perturbation theory at low energies. It shows that the multi-particle cross
sections are likely to be small at all energies, except, maybe, for
exponentially high ones. Consider $\lambda \phi^{4}$ theory for
definiteness (the argument works also for instanton-like processes). The
propagator in this theory obeys the K\"allen--Lehmann relation, which I use
at small $Q$, say $Q=0$,
\begin{equation}
             G(Q=0) = \int~ds
	\frac{\sigma_{tot}(s)}{s}
\label{9*}
\end{equation}
(the argument
is easily generalized when subtractions are necessary), where
$\sigma_{tot}(s)$ is the total cross section of the "decay" of one virtual
boson with energy $\sqrt{s}$ into arbitrary nubmer of real bosons. The left
hand side of this relation is believed (and in some cases proved) to be
a nice asymptotic series in $\lambda$ whose finite number, $k\ll 1/\lambda$,
of terms are given by perturbation theory. These $k$ terms on the right
hand side come from the cross sections of production of $k$ particles and
less. Since the remaining part of the left hand side is smaller than
$\const\cdot\lambda^{k+1}$, the sum of contributions of $n$-particle cross
sections with $n > k$ into the right hand side is small,
\[
   \sum_{n=k+1}^{\infty}
        \int~ds \frac{\sigma_{n}(s)}{s} < \const \cdot \lambda^{k+1}
\]
This excludes the possibility that the $n$-particle cross sections at
$n\sim 1/\lambda$ are large at energies of order $m/\lambda$. In fact, in
view of Eq. (\ref{8*}), these cross sections are expected to be
exponentially small, i.e., the exponent $F$ is expected to be negative at
any $n$.

This argument does not, however, exclude a still very intersting
possibility that $F$ tends to zero as $\lambda n \to \infty$ and/or
$\epsilon \to \infty$. In any case, theoretical understanding of the
exponential behavior of the cross sections is of considerable importance.

Equation (\ref{9*}) actually provides a possible way to estimate (or rather
put an upper bound on) $n$-particle cross sections \cite{Wiley}. The left
hand side can be non-perturbatively calculated on a lattice and then
compared to a few terms in the perturbative expansion of $G(Q=0)$. The
difference is then attributed to the contributions of $n$-particle cross
sections at large enough $n$. This approach has already
been tested in $(1+1)$-dimensional scalar theories \cite{Wiley}.

\section{Regular Classical Solutions: from {\it many $\to$ many}
            to {\it few} $\to$ {\it many}}

One possible way \cite{RTMany,TMany} to study the probability of the
processes {\it few} $\to$ {\it many} is to consider, as the first step, the
processes {\it many} $\to$ {\it many}, i.e.,
\[
  n_{in} \to n
\]
where $n_{in}$ , the number of incoming particles, is formally of order
$1/\alpha$, the inverse coupling constant. For finite $\alpha n_{in}$
and small $\alpha$, both the initial and final states contain
parametrically large number of particles, and it is natural that the
scattering process can be described in semiclassical terms. The probability
of {\it few} $\to$ {\it many} process may then be obtained in the limit
$\alpha n_{in} \to 0$.

One realization of this idea is to consider real classical solutions to
Minkowskian field equations, i.e., scattering of classical waves. To every
classical solution that disperses into free waves at $t \to \pm \infty$
one can assign the number of incoming and outgoing particles, both of which
are naturally of order $1/\alpha$. The probability of the scattering of
these multiparticle states is not suppressed. At given energy one tries to
minimize the number of incoming particles under the condition that the
topological number changes by 1 (instanton-like transitions) or that the
number of outgoing particles is fixed (processes without instantons). If
the minimum number of incoming particles tends to zero (in units
$1/\alpha$) as the total c.m. energy approaches some $E_{cr}$, then
{\it few} $\to$ {\it many} processes are not suppressed exponentially at
$E>E_{cr}$ (this includes more likely possibility that $E_{cr}=\infty$,
in which case the exponential suppression of {\it few} $\to$ {\it many}
cross sections disappears at asymptotically high energies). In the opposite
case when the minimum number of incoming particles  needed to induce the
classical transition remains larger than
$\const \cdot 1/\alpha$, the exponential
suppression of {\it few} $\to$ {\it many} persists at all energies, but the
actual exponent cannot be calculated by studying classical scattering.

Naturally, obtaining general enough set of analytical classical
solutions is possible only in a very narrow set of models. In specially
designed models \cite{RubSon} (false vacuum decay in scalar theories with
potentials of exponential type  in $(1+1)$ dimensions) it has been found
that the number of incoming particles required for the classical
instanton-like transition to occur is finite in units $1/\alpha$ at all
energies. This provides an example of
{\it few} $\to$ {\it many} transitions
whose rate is exponentially small at arbitrarily high  energies.

In realistic models, this program requires extensive numerical
calculations. Most impressive results up to now have been obtained in ref.
\cite{Rebbi} where $(3+1)$-dimensional $SU(2)$ Yang--Mills--Higgs theory
has been studied. It has been found that, indeed, the minimum number of
incoming particles producing classically instanton-like (or
rather sphaleron-like) transitions  decreases with energy. Though the
results of ref. \cite{Rebbi} are not decisive yet (in these calculations
$n_{in}$ dropped by about 30 per cent only), this study suggests that the
full scale realization of this program is quite feasible.  A remarkable
feature of this study is that it proved "experimentally" that
distinguishing processes in a trivial vacuum and sphaleron-like
transitions is indeed possible, and one is really able to study the latter
transitions even though topologically trivial processes may be much more
numerous.

When the number of incoming particles and total energy are such that the
rate is exponentially suppressed, relevant classical Minkowskian solutions
are merely absent.  In that case one is able, however, to
formulate a classical boundary value problem in complex time \cite{RST}.
To fix the number of incoming particles and total energy, one introduces
the Legendre conjugate real parameters, $T$ and $\theta$. The contour in
complex time plane on which the classiacal field equations are solved has
to start at $T=t'+ iT/2$, where $t'=real \to \-infty$. At this asymptotics
the boundary condition for  (spatial Fourier transform of) the field is
\begin{equation}
\phi_{{\bf k}} (t') =
		 b_{{\bf k}}^{*}\e^{i\omega_{k}t'} +
             \e^{-\theta}b_{-{\bf k}}\e^{-i\omega_{k}t'}
\label{14*}
\end{equation}
where $b_{{\bf k}}$ are arbitrary complex functions of ${\bf k}$.
The contour should end at real $t\to\infty$, and another boundary condition
is that the field is real in the future asymptotics. The number of the
boundary conditions is sufficient for determiming the classical solution,
up to translations in space-time. The total probability of the
instanton-like transitions indeed has the exponential form, provided that
$\alpha E$ amd $\alpha n_{in}$ are fixed as $\alpha\to 0$,
\[
   \sigma_{tot} \propto
	       \exp\left[\frac{1}{\alpha}F(\alpha E,\alpha n_{in})\right]
\]
The solution to the above boundary value problem determines the exponent,
\[
     \frac{1}{\alpha}F(\alpha E, \alpha N_{in})
                    = ET -n\theta - 2S_{cl}
\]
where $S_{cl}$ is the real part of the classical action with Euclidean
time convention, and $T$ and $\theta$ are related to $E$ and $n_{in}$ by
\[
	E = 2\frac{\partial S}{\partial T},~~~~
	n_{in} = 2\frac{\partial S}{\partial \theta}
\]
Thus, the study of {\it many $\to$ many} transitions in the classically
forbidden region also reduces to the problem of solving classical field
equations, but now for {\it complex} classical fields in {\it complex} time
plane. Analytical results are again available only in specially designed
models \cite{SonRub} where they nicely match those obtained by studying the
classical scattering \cite{RubSon}. There has been very few attempts to
find the corresponding classical solutions numerically \cite{Rebbi,KuTi}.
The most intersting results up to now have been obtained in ref. \cite{KuTi}
where false vacuum decay in $\phi^4$ theory has been studied in $(3+1)$
dimensions. The energy range accessible was from $0.8 E_{sph}$ to
$3 E_{sph}$; it was found that at the highest energy in this interval,
the exponential suppression disappears at $n_{in}=0.4 n_{sph}$, where
$E_{sph}$ and $n_{sph}$ are energy and number of particles charactristic to
the sphaleron decay, $E_{sph}$,$n_{sph} \sim 1/\alpha$. The results of ref.
\cite{KuTi} indicate that the exponential suppression does not disappear at
high energies at $n_{in} \ll n_{sph}$, and this expectation should be
confirmed or disproved in near future.

Thus, the two approaches which make use of the idea of {\it many $\to$
many} transitions complement each other and are capable to provide a
coherent picture of the instanton-like processes at high energies. The
existing results indicate exponential suppression of {\it few} $\to$
{\it many} processes, and decisive results are expected to be obtained soon.

\section{Singular Classical Solutions: towards the Generalization of
Landau Technique to Quantum Field Theory}

It is well known in quantum mechanics of one variable that semiclassical
matrix elements can be evaluated by making use of singular classical
solutions to equations of motion in Euclidean time \cite{Landau}. Namely,
one considers matrix elements of the type
\begin{equation}
   \bra{E_{2}}\hat{O}\ket{E_{1}}
\label{17*}
\end{equation}
where $\ket{E_{1}}$ and $\ket{E_{2}}$ are highly excited (semiclassical)
states, and $\hat{O}$ is an operator independent of $\hbar$. The Landau
technique states that this matrix element is exponentially small,
\[
   \bra{E_{2}}\hat{O}\ket{E_{1}} \propto \e^{-S/\hbar}
\]
and the following properties hold: i) the exponential factor is independent
of the operator $\hat{O}$; ii) The exponent $S$ is equal to the truncated
action of a singular classical solution in Euclidean time, which has
energies $E_{1}$ and $E_{2}$ before and after the singularity,
respectively; this solution begins and ends at classical turning points.

It has been pointed out by several authors
\cite{VolLand,Khlebnikov,DiakPetLandau,VolGor,CornTik}
that there exists a remarkable similarity between the matrix elements
(\ref{17*}) and amplitudes of multi-particle processes at high energies, so
the Landau technique may be generalizable to field theory. In fact, this
technique was successfully used for the evaluation of tree amplitudes at
threshold in $\phi^4$ theory \cite{VolLand} and also for estimating full
amplitudes at threshold in this theory at asymptotically large number of
final particles \cite{VolGor}.

Indeed, consider, as an example, multi-particle production in
$\frac{\lambda}{4}\phi^{4}$ theory (no instantons). The corresponding
amplitudes are
\begin{equation}
	   \bra{n}\hat{O}\ket{0}
\label{18*}
\end{equation}
where both vacuum and final states may be viewed as
semiclassical states at
$n\sim\frac{1}{\lambda}$, and
$\hat{O}$ is $\phi (x)$ or
$\phi (x) \phi (y)$
(for one-particle or two-particle initial states,
respectively);
$\hat{O}$ obviously does not depend on $\lambda$. The similarity
between Eqs. (\ref{17*}) and (\ref{18*}) is clear. Further support to the
idea to generalize the Landau technique comes from perturbative
calculations \cite{LibSonTro} which show that the exponent for the
amplitude near threshold is in fact independent of the operator $\hat{O}$
in Eq. (\ref{18*}), at least at exponentiated one loop order for operators
like
$\phi (x)$ and
$\phi (x) \phi (y)$. Most notably, singular solutions appear naturally in
the calculations of {\it tree} cross sections at and above threshold.

\subsection{Singular Solutions and Tree Cross Sections}

Let me discuss in more detail the relevance of singular classical solutions
to the tree cross sections. I am going to show that the tree  $n$-particle
cross sections at $\lambda \to 0$ with $\lambda n, E/n =$fixed, are related
to singular solutions to Euclidean field equations. The resulting
prescription has been suggested in ref. \cite{Son} on slightly different
grounds.

To begin with, consider generating function for $1 \to n $
tree amplitudes
{\it at $n$-particle threshold}. It has been shown \cite{Brown} that
it obeys Minkowskian field equation
\[
  \partial^{2}\phi + m^{2} \phi + \lambda \phi^{3} = 0
\]
and it is homogeneous in space, $\phi = \phi (t)$. The relevant solution is
\begin{equation}
   \phi(t) =
    \frac{z_{0}\e^{imt}}{1 -
    \left(\frac{\lambda}{8m^2}\right)z_{0}^{2}\e^{2imt}}
\label{19*}
\end{equation}
where $z_{0}$ is a free parameter. The tree amplitudes at threshold are
obtained by writing this solution in the following form,
\[
   \phi(z(t)) =
    \frac{z(t)}{1 -
    \left(\frac{\lambda}{8m^2}\right)z(t)^{2}}
\]
and expanding in $z(t)$, i.e.,
\[
    \bra{n, {\bf p_{1}}=\dots={\bf p_{n}} = 0}\phi\ket{0}^{tree} =
	      \left[\frac{d^{n}}{dz^{n}}\phi(z)\right]_{z=0}
\]
In this way one recovers the amplitudes (\ref{7*}).

Being continued to complex time, the solution (\ref{19*}) decays at
$Im~t \to\infty$, and has a singularity at
\[
       t_{0}= i\tau_{0} =
       \frac{i}{2m}\ln\frac{8m^2}{\lambda z_{0}^{2}}
\]
Viewed as the field configuration in Euclidean time, this solution has flat
surface of singularities near which
\begin{equation}
   \phi = \sqrt{\frac{2}{\lambda}}\frac{m}{l}
\label{19+}
\end{equation}
where $l={\em Im}~t - \tau_{0} = \tau - \tau_{0}$ is the distance to this
surface. The flatness (independence of ${\bf x}$) of this surface can be
viewed as a reflection of the fact that all spatial momenta of outgoing and
incoming particles are zero.

This way the singular solutions emerge may appear accidental. To see
that this is not so, consider $1\to n$ process at the tree level above
$n$-particle threshold. Clearly, the amplitudes themselves may depend on
the correlations between the momenta of the outgoing particles, so the form
of the amplitude is not expected to be simple. On the other hand, the total
(integrated over phase space) $n$-particle cross section depends only on
$n$ and total energy $E$. It is therefore natural to look for a
semiclassical way of calculating this cross section.

As the first step, let me consider the amplitude of the production of a
coherent state
$\ket{\{b({\bf k})\}}$
at given energy $E$. It is given by
the functional integral in the coherent state representation with
appropriate boundary terms at $t=T_{f}\to\infty$,
 \begin{equation}
\bra{\{b({\bf k})\}}\tilde{\phi}(E)\ket{0} =
	       \int~D\phi ~\e^{iS + B(b,\phi_{f})}~\tilde{\phi}(E)
\label{20*}
\end{equation}
where $\phi_{f} =  \phi(T_{f})$,
\begin{equation}
   B=
   -\frac{1}{2} \int ~d{\bf k}~ b^{*}({\bf k}) b^{*}(-{\bf k})
   -\frac{1}{2}
   \int ~d{\bf k}~\omega_{k}\phi_{f}({\bf k})\phi_{f}(-{\bf k}) +
  \int ~d{\bf k}~\sqrt{2\omega_{k}} b^{*}({\bf k}) \phi_{f}(-{\bf k})
\label{20+}
\end{equation} and
\[
\tilde{\phi}(E) = \int~d{\bf x}dt~
   \phi(x) \e^{-iEt + i{\bf Px}}
\]
Here $E\sim 1/\lambda$ is the total
c.m. energy, and ${\bf P} =0$ is the total c.m. momentum. The  amplitude
(\ref{20*}) is determined {\it at the tree level} by the saddle point of
the exponent, which is a solution to the classical equation \[
  \partial^{2}\phi_{c} + m^{2} \phi_{c} + \lambda \phi^{3}_{c} = 0
\]
The boundary conditions for $\phi_{c}$ are obtained by
varying the fields at
$t=T_{f}\to+\infty$ and $t=T_{i}\to -\infty$. One finds that $\phi_{c}$ has
only positive-frequency part at $t\to -\infty$ (Feynman boundary
conditions),
\begin{equation}
 \phi_{c}({\bf k},t)
		     \propto \e^{i\omega_{k}t},
			   ~~~~~ t\to -\infty
\label{21*}
\end{equation}
and at $t \to + \infty$ its positive-frequency part is determined by
$b^{*}({\bf k})$,
\begin{equation}
    \phi_{c}({\bf k}, t) =
	     \frac{b^{*}({\bf k})}{\sqrt{2\omega_{k}}}\e^{i\omega_{k}t}
	     + (arbitrary)\cdot \e^{-i\omega_{k}t},
		   ~~~~~t \to +\infty
\label{21**}
\end{equation}
In fact, the energy of the field (\ref{21*}) is zero, so $\phi_{c}$ has
only positive-frequency part at $t\to +\infty$ by energy conservation,
\begin{equation}
    \phi_{c}({\bf k}, t) =
	     \frac{b^{*}({\bf k})}{\sqrt{2\omega_{k}}}\e^{i\omega_{k}t}
		   ~~~~~t \to +\infty
\label{21+}
\end{equation}
The exponent in Eq. (\ref{20*})  turns out to be zero for $\phi =
\phi_{c}$, so the tree amplitude has particularly simple form,
\begin{equation}
\bra{\{b({\bf k})\}}\tilde{\phi}(E)\ket{0} =
		 A_{E}(b^{*}) =
               \int~d{\bf x}dt~\phi_{c}(b^{*};x)\e^{-iEt + i{\bf Px}}
\label{22*}
\end{equation}

The second step is to extract the $n$-particle cross section from the
amplitude (\ref{22*}). To do this, one notes that, in the spirit of ref.
\cite{RTMany}, the generating function for $n$-particle cross sections has
the following form,
\begin{equation}
     \Sigma(\xi;E) =
    \frac{1}{F} \int~DbDb^{*}
    \exp \left(\int~d{\bf k}~b({\bf k})b^{*}({\bf k})\right)
    A_{E}(\sqrt{\xi}b^{*})
    \bar{A}_{E}(\sqrt{\xi}b)
\label{22**}
\end{equation}
where $F$ is the usual flux factor. Indeed, the $n$-particle amplitude is
proportional to the term containing a factor
$b^{*}({\bf k_1})\dots b^{*}({\bf k_n})$  in the expansion of
$A_{E}(b^{*})$ in $b^{*}$'s. The expansion of Eq. (\ref{22**}) in $\xi$
is thus the expansion in the number of outgoing particles, and integration
over
$b$, $b^{*}$
performs, as usual, the integration over phase space.
Thus one obtains, after a change of variables,
\begin{equation}
   \sigma_{1\to n}^{tree}(E) =
    \frac{1}{F} \int~\frac{d\xi}{\xi^{n+1}}~DbDb^{*}
    \exp \left(-\frac{1}{\xi}\int~d{\bf k}~b({\bf k})b^{*}({\bf k})\right)
    A_{E}(b^{*})
    \bar{A}_{E}(b)
\label{22+}
\end{equation}
where the integration contour in complex $\xi$ plane surrounds the origin.

I am going to perform the integration in Eq. (\ref{22+}) in the saddle
point approximation, which is valid at $n\sim 1/\lambda$. However, the
integration over
$b$, $b^{*}$
contains zero mode due to space-time translations. Also,
the expression (\ref{22*}) contains exponentially large factor due to large
$E$ ($E\sim 1/\lambda$). Indeed, let me parametrize
\[
    b^{*}({\bf k}) =
	     \beta^{*}({\bf k}) \e^{i\omega_{k}t_{0} - i{\bf kx}}
\]
and assume that the set $\{\beta^{*}({\bf k})\}$ obeys four constraints
(about which I will have to say more later). Then, due to the boundary
conditions (\ref{21*}), (\ref{21**})
\[
   \phi_{c}(b^{*};x) =
   \phi_{c}(\beta^{*};x +x_{0} )
\]
so that
\[
    A_{E}(b^{*}) =
	 \e^{iEt_{0} - i{\bf Px_{0}}} A_{E}(\beta^{*})
\]
The integration over
$b$, $b^{*}$
becomes the integration over
$\beta$, $\beta^{*}$ and $x_{0}$, $x'_{0}$, i.e.,
\[
   \sigma_{1\to n}^{tree}(E) =
    \frac{1}{F} \int~\frac{d\xi}{\xi}~D\beta D\beta^{*}~
    dt_{0}dt'_{0}~ d{\bf x_{0}} d{\bf x'_{0}}
\]
\[
    \exp\left[-n\ln \xi + iE(t_{0} - t'_{0})
    -i{\bf P}({\bf x_{0}} - {\bf x'_{0}})\right]
\]
\[
    \exp \left[-\frac{1}{\xi}\int~d{\bf k}~
    \beta({\bf k})\beta^{*}({\bf k})
    \e^{i\omega_{k}(t_{0} - t'_{0}) - i{\bf k}({\bf x_{0}} - {\bf
    x'_{0}})}\right]
           J A_{E}(\beta^{*}) \bar{A}_{E}(\beta)
\]
where $J$ contains $\delta$-functions of the constraints and the
corresponding Faddeev--Popov determinant.

Let me now  assume that the constraints are chosen in such a way that
$A_{E}(\beta^{*})$ does not contain exponential factors. Then the integral
over $(x_{0} + x'_{0})$ gives rise to the usual volume factor in the cross
section (which is cancelled by the flux factor $1/F$). The integration over
remaining variables is of saddle point nature. The saddle point value of
$({\bf x_{0}} - {\bf x'_{0}})$ is zero, so I omit this variable in what
follows. Denoting
\begin{equation}
    \xi = \e^{\theta}~~~~~~t_{0} - t'_{0} = -iT
\label{24**}
\end{equation}
one finds the effective "action"
\begin{equation}
   W_{tree}(T,\theta,\beta^{*},\beta) =
	ET - n\theta -
	\e^{-\theta}
	\int~d{\bf k}~
	\beta({\bf k})
	\beta^{*}({\bf k})
	\e^{\omega_{k}T}
\label{24*}
\end{equation}
which should be extremized with respect to $T$, $\theta$, $\beta$ and
$\beta^{*}$ under the constraints to which  I turn now.

The constraints to be imposed on $\beta^{*}({\bf k})$ should fix the
translational invariance in space and time. Let me impose them implicitly by
requiring that $A_{E}(\beta^{*})$ does not contain exponential factors in
spite of the fact that $E$ is of order $1/\lambda$. One expects that
relevant classical solutions are generalizations of Eq. (\ref{19*}). Indeed,
due to Eqs. (\ref{21*}) and (\ref{21+}) the soultions should decay as
$Im~t \to +\infty$. In other words, the solutions considered in
Euclidean space-time should vanish as $\tau \to +\infty$, where $\tau$ is
the Euclidean time coordinate. Since there are no instantons, the
solutions must be singular in Euclidean space-time. Generically, the points
of singularities form a three-dimensional surface $\tau=\tau({\bf x})$,
near which the field behaves like
\begin{equation}
   \phi_{c}\sim\sqrt{\frac{2}{\lambda}}\frac{1}{l}
\label{25*}
\end{equation}
where $l$ is the distance to this surface (cf. Eq. (\ref{19+})). The right
hand side of Eq. (\ref{22*}) for such a solution behaves like
$\exp(E\tau_{m} + i {\bf Px_{m}})$ where $\tau_{m}$ and ${\bf x_{m}}$ are
the Euclidean coordinates of the singular point closest to the Minkowski
time axis ($\tau_{m}< 0$ for the solution be smooth  everywhere in
Minkowski space-time). For the right hand side of Eq. (\ref{22*}) to
contain no exponential factors, one requires $\tau_{m} \to 0$,
${\bf x_{m}} = 0$. In other words, the singularity surface in Euclidean
space-time should touch from below the surface $\tau = 0$ at the origin.
This constraint fixes the translational invariance (both in Minkowskian and
Euclidean space-time).

So, the exponent of the tree cross section is obtained in the following
way. One considers Euclidean classical solutions $\phi_{c}({\bf x},\tau)$
which decay at $\tau \to +\infty$ and have singularities on surfaces
$\tau=\tau_{s}({\bf x})$. For any given singularity surface there exists at
most discrete set of such solutions: indeed, one may first consider a well
defined boundary  value problem $[~\phi \to 0$ at $\tau \to +\infty$;
$\phi\left(\tau = \tau_{s}({\bf x})\right) = \Lambda]$
and then send $\Lambda \to
\infty$. The behavior of the solution at large $\tau$,
\begin{equation}
   \phi({\bf k},\tau) =
   \frac{\beta^{*}({\bf k})}{\sqrt{2\omega_{k}}}\e^{-\omega_{k}\tau}
\label{26*}
\end{equation}
determines the functions $\beta^{*}({\bf k})$ for a given surface
$\tau_{s}$. Then one finds an extremum of the right hand side of
Eq. (\ref{24*})  among all surfaces obeying the constraint
\[
  \tau_{s}({\bf x} =0) =0
\]
and having $\tau_{s}({\bf x} \neq 0)<0$. Finally, the extremum with respect
to $T$ and $\theta$ of the resulting $W(T,\theta)$ determines $T$ and
$\theta$ in terms of $E$ and $n$, and the cross section is
\[
   \sigma^{tree}_{1\to n} \propto
   \exp \left( W_{tree}[T(E,n),\theta(E,n)]\right)
\]
This prescription coincides with one advocated in ref. \cite{Son}.

The dependence on $\lambda$ of the tree $n$-particle cross section is, in
fact, obvious from Feynman graphs,
$\sigma^{tree}_{1 \to n} \propto \lambda^{n}$. This dependence is easily
restored within the above prescription by the following scaling argument.
Since the integral
\begin{equation}
	\int~d{\bf k}~
	\beta({\bf k})
	\beta^{*}({\bf k})
	\e^{\omega_{k}T}
\label{27*}
\end{equation}
as well as the constraints on the surface of singularities, do not depend
on $\theta$, the extremum value of this integral depends only on $T$,
\[
	\int~d{\bf k}~
	\beta({\bf k})
	\beta^{*}({\bf k})
	\e^{\omega_{k}T} = \frac{1}{\lambda} N(T)
\]
(the dependence on $\lambda$ comes about from even more simple scaling),
where $N(T)$ should be determined by solving the classical problem oulined
above. Therefore, the functional form of $W_{tree}$ is
\[
    W_{tree}(T,\theta) =
	ET - n\theta - \e^{-\theta}\frac{1}{\lambda} N(T)
\]
Then the extremum with respect to $\theta$ is at
$ \theta = - \ln \frac{\lambda n}{N(T)} $
and the value of $W_{tree}$ at this extremum is
\begin{equation}
   W_{tree}(T) = ET + n\ln (\lambda n) - n \ln N(T)
\label{28*}
\end{equation}
After extremizing this expression with respect to $T$, one finds the
functional form of $W_{tree}(E,n)$,
\[
W_{tree}(E,n) = n\ln (\lambda n) + n\Psi (\epsilon)
\]
where, as before, $\epsilon = E/n$ is the average energy per an outgoing
particle. Therefore, the functional form of the tree cross section is
\begin{equation}
   \sigma^{tree}_{1 \to n} \propto \e^{W(E,n)} \propto
            n!\lambda^{n}\e^{n\Psi(\epsilon)}
\label{28+}
\end{equation}
which has correct dependence on $\lambda$. Note that the tree cross section
increases with $n$ at given $\epsilon$, i.e., it indeed hits the unitarity
bound at large enough $n$.

Let me also point out that the extremum of the integral (\ref{27*}) is
naturally expected to be a minimum; the perturbative calculations suggest
that the extremum of $W(T)$, Eq. (\ref{28*}), {\it is also a minimum}.
[The latter observation is in accord with the fact that the integration
over the original variable $(t_{0} - t'_{0})$ runs along the Minkowski time
axis, and the extremum of $W$ is a maximum along this line. Switching to
$T$ according to Eq. (\ref{24**}) converts the extremum into a minimum.]
Therefore, it is possible to formulate a
variational procedure for obtaining
the lower bound on $\Psi(\epsilon)$ entering Eq. (\ref{28+}); this
procedure has been applied to $\phi^{4}$ theory in ref. \cite{Bezrukov}.

Thus, there exists a well defined procedure that relates the exponent for
the tree $n$-particle cross section to singular solutions of classical
field equations in Euclidean space-time. The existing perturbative results
have been
reproduced within this approach \cite{Son}, and the
evaluation of the behavior of tree cross sections at
all energies becomes a
matter of numerical calculations.

{\it Loops}

As discussed in sect. 2.1, loop effects become important at
$\lambda n \sim 1$, and the correct behavior of the $n$-particle cross
section is determined by diagrams with arbitrary number of loops. One may
argue, however, that the complete exponent for the cross section, including
loop  effects, is again related to singular classical solutions of  the
field equations. One argument is close in spirit to the approach outlined
in sect.2.2. Consider, instead of the matrix element (\ref{20*}), the
following matrix element,
\begin{equation}
  A(j) =
     \bra{\{b({\bf k})\}}
     P_{E} \exp\left(\int~dx~ j(x)\phi(x)\right)
     \ket{0}
\label{30*}
\end{equation}
where $P_{E}$ is the projector onto the subspace of total energy $E$, and
\[
    j(x) = j_{0}\delta (x)
\]
is an external source. The projection by $P_{E}$ may be written in the
integral form \cite{KRTPeriodic}, so that
\[
    A(j) = \int~ d\eta ~e^{iE\eta}
     \bra{\{ b({\bf k})e^{i\omega_{k}\eta}\} } \e^{j_{0}\phi(0)}
     \ket{0}
\]
and the matrix element (\ref{30*}) has the functional integral
representation similar to Eq. (\ref{20*}). It is  then straightforward to
write the double functional integral representation for the "cross section"
of the creation of $n$ particles by the external source $j$; in particular,
one makes use of the trick (\ref{22**}). The form of this double functional
integral is, roughly speaking,
\[
   \sigma_{j \to n} =
      \int~D\phi D\phi'~Db Db^{*}~d\eta d\xi
\]
\begin{equation}
   \exp\left( iS[\phi] - iS[\phi'] + j_{0}\phi(0)
      + j_{0}\phi'(0) + iE\eta +n\ln \xi + \dots \right)
\label{31*}
\end{equation}
where dots stand for boundary terms like (\ref{20+}) and terms proportional
to $bb^{*}$; all these terms are bilinear in $(b,b^{*},\phi,\phi')$. It is
clear that at
\[
     j_{0}\sim \frac{1}{\lambda}
\]
the integral (\ref{31*}) is of saddle point nature, and its value is
determined by a solution to the classical field equations in the presence
of the source, which obeys the boundary conditions similar to
Eqs. (\ref{21*}) and (\ref{21**}). The "cross section" has then the
following functional form,
\begin{equation}
       \sigma_{j \to n} \propto
		   \exp \left[ n
		   F\left( \lambda n, \frac{E}{n}, \lambda j_{0}\right)
		   \right]
 \label{31+}
 \end{equation}
The amplitudes
{\it few} $\to n$
are obtained from the matrix element
(\ref{30*}) when $j_{0}$ is small. Therefore, one argues that the exponent
of the full cross section
{\it few} $\to n$ is obtained by taking the limit $\lambda j_{0} \to 0$
in $F\left( \lambda n, E/n, \lambda j_{0}\right)$. In this limit the saddle
point configuration obeys sourceless classical field equation. Clearly,
this configuration must be singular,  otherwise the energy of this
configuration would be the same at $t \to -\infty$ and $t \to +\infty$,
while these energies should be equal to $0$ and $E$, respectively. The
nature of the singularity is to be determined by the above limiting
process.

This idea has been elaborated in ref. \cite{Son}, where the classical
boundary value problem for fields with subsequent extremisation over their
singularity surfaces has been formulated. In the limit of small $\lambda
n$, when the tree cross section is a good approximation, this procedure
reduces to one discussed above. The exponentiated one loop correction has
been also reproduced by this technique \cite{Son}. So, {\it known results
summarized in Eq. (\ref{8**}) indeed follow from the semiclassical
treatment that generalizes the Landau technique to field theory}. It
remains to be understood what  kind of singularity of the
calssical field is relevant in the general case, and whether this technique
is suitable for numerical studies.

\section{Summary}

To summarize, {\it few} $\to$ {\it many} cross sections cannot, in general,
be understood within perturbation theory, irrespectively of whether one
considers instanton-like processes or processes in the trivial vacuum
sector. The functional form of the cross sections in both cases strongly
suggests the possibility of semiclassical treatment of these processes.
Several semiclassical approaches have been recently developed and tested
analytically and numerically. The decisive results have not been
obtained yet, but they are expected to appear in near future.

This work is supported
in part by INTAS grant 94-2352  and
International Science Foundation grant MKT300.

\end{document}